\documentclass[aps,prc,preprint,groupedaddress,superscriptaddress]{revtex4-1}
\usepackage[utf8]{inputenc}
\usepackage{graphicx}
\usepackage{amsmath, amssymb, bm, mathrsfs, gensymb, longtable}

\begin{document} 

\title{Isospin mixing and Coulomb mixing in ground states of even-even 
nuclei}

\author{Bui Minh Loc} 
\email[]{minhlocbui@mail.tau.ac.il}
\affiliation{School of Physics and Astronomy, Tel Aviv University, Tel Aviv 
69978, Israel.}
\author{Naftali Auerbach} 
\affiliation{School of Physics and Astronomy, Tel Aviv University, Tel Aviv 
69978, Israel.}
\author{G. Col\`{o}} 
\affiliation{INFN, sezione di Milano, via Celoria 16, I-20133 Milano, Italy}
\affiliation{Dipartimento di Fisica, Università degli Studi di Milano, via 
Celoria 16, I-20133 Milano, Italy}

\date{\today}

\begin{abstract}
In this work, the Coulomb mixing and the isospin mixing in the ground states of 
even-even nuclei are evaluated in perturbation theory. The calculation of the 
isospin mixing is performed by using the connection to isovector monopole 
resonance properties. The uncertainty in the results that depends on different 
choices of the Skyrme interactions is shown. While Coulomb mixing turns out to 
be large in the ground states of heavy nuclei, isospin mixing is very small.
\end{abstract}



\maketitle

\section{Introduction}
The best-known part of the nuclear Hamiltonian is the Coulomb interaction 
between protons $V_C$. As a consequence isospin breaking is dominated by the 
$V_C$. The parent state $| \pi \rangle$ of the nucleus with isospin $T$ and $T_z 
= T$ contains the admixtures of states with isospin $T+1$,
\begin{equation}
 | \pi \rangle = (1 - \sum_\alpha \varepsilon_\alpha^2 )^{1/2}|T, T; 0
\rangle + \sum_\alpha \varepsilon_\alpha|T + 1, T; \alpha \rangle.
\end{equation}
The total probability $\varepsilon^2 = \sum_\alpha \varepsilon_\alpha^2$ is 
\textit{the isospin mixing}. In first-order perturbation theory, the 
expression for isospin mixing is defined as
\begin{equation}\label{eps2tp1}
\varepsilon^2_{T+1} = \sum_{\alpha \neq 0} \frac{|\langle T, T; 0 | 
V^{\rm (IV)}_C | T+1, T; \alpha \rangle|^2}{(E_\alpha - E_{0})^2},
\end{equation}
where $|T, T; 0 \rangle$ denotes the g.s. at the energy $E_{0}$, and 
$\alpha$ are the various quantum numbers needed to specify the states $| 
\alpha \rangle$ at their energy $E_\alpha$. The Coulomb interaction can be 
rewritten in terms of isoscalar, isovector, and isotensor parts, but only the 
isovector part $V^{\rm (IV)}_C$ is kept because the isoscalar part does not 
contribute to (\ref{eps2tp1}) and the isotensor part is small because of the 
long-range nature of the Coulomb interaction. Note that one needs to indicate $T 
+ 1$ because an isovector operator excites not only $|T + 1, T; \alpha 
\rangle$, but also $|T, T; \alpha \rangle$. If the $|T, T; \alpha \rangle$ are 
also taken into account, the admixture is usually much larger. This kind of 
mixing was defined as the Coulomb mixing \cite{AUERBACH1983273}: 
\begin{equation}\label{eps2C}
\varepsilon^2_{C} = \sum_{\alpha \neq 0} \frac{|\langle 0 | V_{C}^{\rm (IV)} | 
 \alpha \rangle |^2}{(E_\alpha - E_{0})^2},
\end{equation}
where the states $|\alpha \rangle$ now include both $T$ and $T+1$ 
excitations. Therefore, \textit{the Coulomb mixing represents the total change 
induced by the Coulomb force in the wavefunction of the ground state}. This 
mixing is more general than just isospin mixing because in many instances the 
effects of the Coulomb force do not lead necessarily to large components that 
differ in the isospin quantum numbers. In nuclei with $N = Z$, the 
isospin mixing and the Coulomb mixing are the same, $\varepsilon^2_{C} = 
\varepsilon^2_{T+1}$, because there are only the $T = 1$ states. In a $N > Z$ 
nucleus, $\varepsilon^2_{C} > \varepsilon^2_{T+1}$ as now there are $|T, T; 
\alpha \rangle$ and $|T+1, T; \alpha \rangle$ contributing.

In Ref.~\cite{AUERBACH1983273} and references therein, it was shown how the 
isospin mixing is connected to the notion of the isovector monopole (IVM) 
resonance that is defined by the operator
\begin{equation}\label{QIV0}
 Q^{\rm (IV)}_0 = \sum_{i} r_i^2 t_z(i),
\end{equation}
where $t_z$ is the $z$-component of isospin operator. Attempts were made to 
observe the IVM resonance experimentally \cite{PhysRevC.34.1822, 
PhysRevLett.118.172501} because it plays an important role in many isospin 
processes \cite{AUERBACH1983273}. Isospin mixing, that is the non-conservation 
of isospin quantum number is a good example. While in the past, the calculations 
\cite{AUERBACH1983273} were performed with a few interactions such as SIII, SIV 
\cite{BEINER197529}, nowadays, there are many Skyrme parameter sets. It is 
useful to study the dependence of the value of the isospin mixing and also the 
IVM properties with different modern Skyrme interactions.

The next section describes the method of calculation. By connecting the isospin 
mixing to the IVM resonance, the calculation of isospin mixing can be 
improved when the isospin properties of the IVM operator \cite{AUERBACH198377} 
are used. In the results, we present 12 Skyrme parameter sets including 
SIII \cite{BEINER197529}, SGII \cite{VANGIAI1981379}, SKM* \cite{BRACK1985275}, 
SkP, SkI2 \cite{DOBACZEWSKI1984103}, SLy4 \cite{CHABANAT1998231}, SkO, SkO' 
\cite{PhysRevC.60.014316}, LNS \cite{PhysRevC.73.014313}, SK255 
\cite{PhysRevC.68.031304}, BSk17 \cite{PhysRevLett.102.152503}, and SAMi0 
\cite{PhysRevC.86.031306}. It demonstrates how much the calculation depends on 
the choice of different Skyrme interactions. We restrict our discussion to 
even-even nuclei. The isospin mixing and Coulomb mixing were calculated for $N = 
Z$ including $^{40}$Ca, $^{56}$Ni, and $^{100}$Sn, and $N > Z$ nuclei including 
$^{48}$Ca, $^{78}$Ni, $^{90}$Zr, $^{120}$Sn, and $^{208}$Pb. 

\section{Method of calculation}\label{method}
The two-body Coulomb potential is given as
\begin{equation}\label{VC}
 V_C = \frac{1}{2} \sum_{i,j}^A \frac{e^2}{|\bm r_i - \bm r_j|} 
\left(\frac{1}{2} - t_z(i) \right) \left(\frac{1}{2} - t_z(j) \right),
\end{equation}
where $t_z$ is the $z$-component of the nucleon-isospin operator, and its 
eigenvalue is $+\frac{1}{2}$ for neutron and $-\frac{1}{2}$ for proton. In the 
Hartree-Fock (HF) calculation, one has the one-body Coulomb potential that 
can be used in Eq. (\ref{eps2tp1}) and (\ref{eps2C}). A simplification 
is to approximate $V_C^{IV}$ using a homogeneous density distribution,
\begin{equation}
 V^{\rm (IV)}_C = - \frac{Ze^2}{2R^3} \sum_{i=1}^{A}(3R^2 - r_i^2)t_z(i),
\end{equation}
for $r \leq R$. In this case, the isospin mixing becomes
\begin{equation}\label{eps2tp1_3}
\varepsilon^2_{T+1} = \left( \frac{Ze^2}{2R^3} \right)^2 \sum_{\alpha \neq 
0}  \frac{|\langle 0 | Q^{\rm (IV)}_0 | T + 1; \alpha \rangle|^2}
{(E_\alpha - E_0)^2},
\end{equation}
where $ Q^{\rm (IV)}_0$ is the $z$-component of the IVM operator and $R = r_0 
A^{1/3}$. It is pointed out that in this approximation, the results are, of 
course, affected by the choice of the value $r_0$. 

\begin{figure}[t!]
\centering
\includegraphics[scale=0.55]{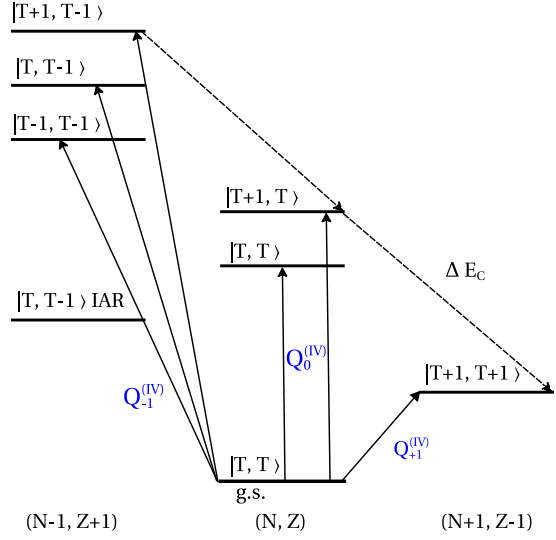}
\caption{Isovector states in nuclei with $N > Z$. A single-particle isovector 
operator $Q^{\rm (IV)}_\mu$, has three components, $Q^{\rm (IV)}_{-1}$, $Q^{\rm 
(IV)}_0$, and $Q^{\rm (IV)}_{+1}$.
In the parent nucleus, $Q^{\rm (IV)}_0$ excites $|T + 1, T; \alpha \rangle$ 
and $|T, T; \alpha \rangle$. In the analog nucleus ($N - 1, Z + 1$), $Q^{\rm 
(IV)}_{-1}$ excites $|T + 1, T - 1; \alpha \rangle$, $|T, T - 1; \alpha 
\rangle$, and $|T - 1, T - 1; \alpha \rangle$. In the nucleus ($N + 1, Z - 1$), 
$Q^{\rm (IV)}_{+1}$ excites $|T + 1, T + 1; \alpha \rangle$ only. 
\label{Fig16BB}}
\end{figure}

In Eq. (\ref{QIV0}), the operator $Q^{\rm (IV)}_0$ is part of a s.p. 
isovector operator $Q^{\rm (IV)}_\mu$ with $\mu = 0, \pm 1$
\begin{equation}
 Q^{\rm (IV)}_\mu = \sum_{i} r_i^2 t_\mu(i),
\end{equation}
where
\begin{equation}
 t_{-1} = +\frac{t_x - it_y}{\sqrt{2}}; \quad t_{+1} = 
-\frac{t_x + it_y}{\sqrt{2}}; \quad \text{and} \quad t_0 = t_z.
\end{equation} 
When the isovector operator $Q^{\rm (IV)}_0$ is applied in the parent nucleus 
with $N > Z$, both $|T, T; \alpha \rangle$ and $|T + 1, T; \alpha \rangle$ are 
excited and the isospin of these states cannot be distinguished when 
performing complicated HF-RPA calculations involving many orbits and thus many 
particle-hole (p-h) states. We should mention here that the states constructed 
of 1p-1h component only do not have good isospin, and in order to have good 
isospin one has to include certain class of 2p-2h components 
\cite{AUERBACH1983273, AUERBACH1979173}. These components are small and usually 
are not included. Their effect on the calculation of isospin mixing as performed 
here is very small.
A technique based on the properties of isovector states in nuclei with $N > Z$ 
(see Fig.~\ref{Fig16BB}) is used to determine separately $|T + 1, T; \alpha 
\rangle$ for the sum in Eq.~(\ref{eps2tp1_3}). First, the calculation using 
operator $Q^{\rm (IV)}_{+1}$
\begin{equation}
 Q^{\rm (IV)}_{+1} = \sum_{i=1}^A r_i^2 t_{+1}(i)
\end{equation} 
that excites only $|T + 1, T + 1; \alpha \rangle$ in the nucleus ($N+1, Z-1$) 
was done. 
After that, $|T + 1, T; \alpha \rangle$ states in the parent nucleus were 
obtained by using the fact that their energies $E_0(T+1; \alpha)$ differ from 
the energies of $|T + 1, T + 1; \alpha \rangle$ in the nucleus ($N + 1, Z - 1$) 
by one Coulomb displacement energy (CDE), $\Delta E_C$, i.e.
\begin{equation}
 E_0(T+1; \alpha) - E_{+1}(T+1; \alpha) = \Delta E_C.
\end{equation}
We used the notations $E_\mu$ ($\mu = 0, \pm 1$) for the energies in the three 
nuclei in Fig.~\ref{Fig16BB}. The CDE, $\Delta E_C$, can be obtained from the 
Skyrme-HF calculation.
The transition strengths to various isospin components $T'$ of the 
$Q^{\rm (IV)}_\mu$ matrix elements are given by the Wigner-Eckart theorem:
\begin{eqnarray}\label{ST'}
S_{T'}^{(\mu)} (\alpha) &=& |\langle T, T; 0| Q^{\rm (IV)}_\mu | \alpha; T', T 
+ \mu \rangle |^2 \nonumber  \\
&=& |\langle T T 1 \mu | T' T + \mu \rangle |^2 \cdot |\langle T;0|| Q^{\rm 
(IV)} || \alpha; T'\rangle|^2.
\end{eqnarray}
The expression $S_{T'} = \sum_{\alpha}|\langle 0|| Q^{\rm (IV)} || \alpha; 
T'\rangle|^2$ is the total reduced transition strengths. With $\langle T T 1 
0| T +1 \phantom{0} T \rangle^2 = 1/(T+1)$ we find:
\begin{equation}\label{eps2tp1_2}
\varepsilon_{T+1}^2 = \sum_{\alpha \neq 0} \frac{S_{T+1}^{(+1)} 
(\alpha)}{(E_\alpha - E_{0})^2},
\end{equation}
with
\begin{equation}
S_{T+1}^{(+1)} (\alpha) = \frac{1}{T + 1}
 |\langle T;0|| Q^{\rm (IV)} || \alpha; T + 1\rangle|^2,
\end{equation}
and $E_\alpha = E_{0}(T+1; \alpha) = E_{+1}(T+1; \alpha) + \Delta E_C$.

It is useful to recall the isospin properties of the IVM resonance related to 
the calculation performed here. The total transition strength of $Q^{\rm 
(IV)}$ is expressed in terms of three reduced 
transition strengths $S_{T'}$
\begin{equation}\label{mmu0}
 m_\mu(0) = \sum_{T'}\langle T T 1 \mu| T' T + \mu \rangle^2 S_{T'},
\end{equation}
and if $\overline{E}_\mu(T')$ is the centroid energy of states of isospin $T'$ 
excited by $Q^{(\rm IV)}_\mu$, we also have
\begin{equation}\label{mmu1}
 m_\mu(1) = \sum_{T'} \langle T T 1 \mu|T' T+\mu \rangle^2 
S_{T'}\overline{E}_\mu(T').
\end{equation}
Using expression (\ref{mmu0}) and (\ref{mmu1}), we can obtain $S^{(0)}_T$ 
and $E_0(T)$ and determine the isospin energy splitting $\Delta \overline{E}_+$
\begin{equation}\label{DeltaEplus}
 \Delta \overline{E}_+ = \overline{E}_0(T + 1) - \overline{E}_0(T)
\end{equation} 
that relates to the symmetry potential $V_1$ defined by the expression
\begin{equation}\label{potV1}
 V_1 = \frac{A}{T+1} \Delta \overline{E}_+.
\end{equation}

In practice, the sum in Eq.~(\ref{eps2C}) for the calculation of the 
Coulomb mixing was obtained from the HF-RPA code following 
Ref.~\cite{COLO2013142}.
The sum in Eq.~(\ref{eps2tp1}) for the isospin mixing was calculated using the 
HF-RPA code including the charge-exchange mode (HF-pnRPA) \cite{pnRPAcode}.

\section{Results and Discussion}
\begin{figure}[t!]
 \centering
\includegraphics[scale=0.7]{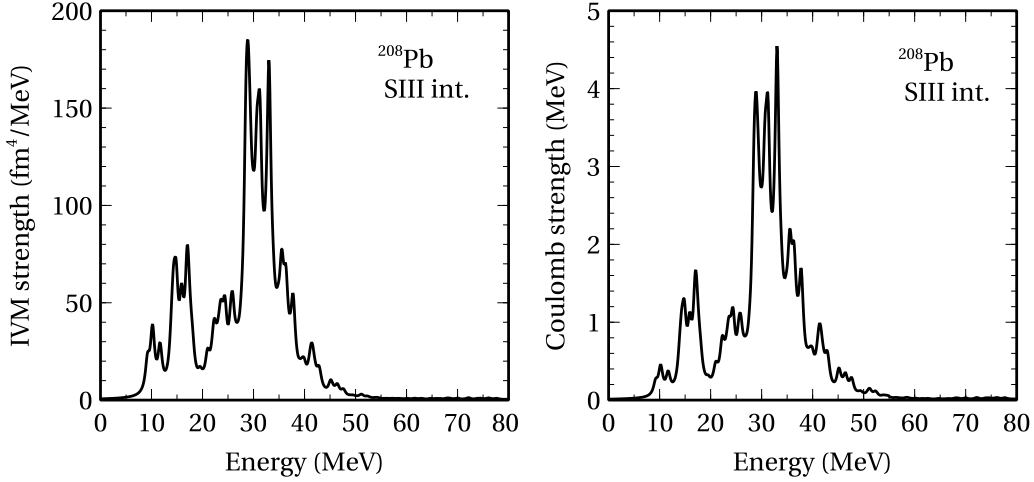}
\caption{The distribution of Coulomb strength and IVM strength in $^{208}$Pb. 
The discrete RPA peaks for both operator were smoothed by using 
the same Lorentzian averaging with the width of 1 MeV. \label{CoulIVM}}
\end{figure}

Although the actual HF Coulomb potential $V_C^{(\rm IV)}$ can be 
used directly as the “probing” operator, in our calculation the IVM operator 
$Q^{(\rm IV)}_0$ is also utilized to have the connection between the Coulomb 
mixing, the isospin mixing and the properties of the IVM resonance. As 
mentioned above, when the IVM operator is used, the results are
affected by the choice of $r_0$. In Fig.~\ref{CoulIVM}, the distribution of 
Coulomb strength evaluated using the $V_C^{(\rm IV)}$ and IVM strength 
evaluated using the $Q^{(\rm IV)}_0$, are shown for $^{208}$Pb. We can see 
in Fig.~\ref{CoulIVM} their close similarity.
This is the reason why one can use the ratio $\eta$ between the total strength 
of Coulomb distribution and that of the IVM distribution:
\begin{equation}
 \eta = \frac{\sum \langle 0| V_C^{\rm (IV)} | \alpha \rangle}{\sum \langle 0| 
Q_0^{\rm (IV)}| \alpha \rangle}
\end{equation}
instead of the factor $\left( \frac{Ze^2}{2R^3} \right)^2$ in 
Eq.~(\ref{eps2tp1_3}).
Therefore, the uncertainty from the value of $r_0$ is avoided. In addition, it 
was found that the ratio $\eta$ is close to the value of the factor in 
Eq.~(\ref{eps2tp1_3}) if $r_0 = 1.25$ fm. In the calculation of 
the Coulomb strength above, the $V_C^{(\rm IV)}$ contains not only the inside 
part ($r \leq R$) but also the outside part ($r > R$). From the similarity 
shown in Fig.~\ref{CoulIVM}, we can conclude that the outside part ($r > R$) 
does not contribute much to the result. 

Table \ref{IsomixNeqZ} shows the values of the Coulomb mixing for $N = Z$ 
nuclei including $^{40}$Ca, $^{56}$Ni, and $^{100}$Sn. In this case, the 
Coulomb mixing and isospin mixing are the same because there are only $T = 1$ 
states.The difference between two different operators is very small, and it 
allows us to use the $Q^{(\rm IV)}_0$ with the ratio $\eta$ instead of $V^{\rm 
(IV)}_C$. The difference in the value of the isospin mixing between different 
Skyrme interactions is not large. 

The Coulomb potential is kept in our HF and RPA calculation.
One can argue that the Coulomb potential should not be included in this 
calculation. The code \cite{COLO2013142} we use allows us easily to include or 
exclude the Coulomb potential consistently (in both HF and RPA). We find that 
this uncertainty, in this case, is not large even in $^{100}_{\phantom{0}50}$Sn 
(see Table \ref{IsomixNeqZ} and Table \ref{IsomixNeqZ2}). 
We prefer to use the results of the HF and RPA included Coulomb potential as 
inputs into the calculation of isospin and Coulomb mixing because they are more 
realistic and can be compared to experiment.

\begin{table}[t!]
\caption{The Coulomb mixing (\%) of $N = Z$ nuclei including $^{40}$Ca, 
$^{56}$Ni, and $^{100}$Sn. The Coulomb potential is \textit{included} in the 
HF-RPA calculation.
\label{IsomixNeqZ}}
 
\centering
\begin{tabular}{|r|l|rr|rr|rr|}
 \hline \hline
& & \multicolumn{2}{c|}{$^{40}$Ca} 
& \multicolumn{2}{c|}{$^{56}$Ni} & 
\multicolumn{2}{c|}{$^{100}$Sn} \\
No. & Int. & $Q^{(\rm IV)}_0$ & $V_C^{(\rm IV)}$ 
& $Q^{(\rm IV)}_0$ & $V_C^{(\rm IV)}$ & 
$Q^{(\rm IV)}_0$ & $V_C^{(\rm IV)}$ \\
\hline
1	&	SIII	&	0.96 	&	0.68 	&	1.55 	&	
1.22 	&	5.44 	&	4.54 	\\
2	&	SGII	&	1.09 	&	0.79 	&	1.85 	&	
1.46 	&	6.58 	&	5.51 	\\
3	&	SKM*	&	1.12 	&	0.78 	&	1.82 	&	
1.42 	&	6.44 	&	5.34 	\\
4	&	SKP	&	1.15 	&	0.81 	&	2.00 	&	
1.56 	&	6.82 	&	5.69 	\\
5	&	SkI2	&	0.87 	&	0.62 	&	1.43 	&	
1.11 	&	5.34 	&	4.46 	\\
6	&	SLy4	&	1.05 	&	0.77 	&	1.78 	&	
1.43 	&	6.17 	&	5.27 	\\
7	&	SKO	&	0.90 	&	0.62 	&	1.31 	&	
0.98 	&	5.04 	&	4.13 	\\
8	&	SKO'	&	1.06 	&	0.73 	&	1.45 	&	
1.13 	&	5.64 	&	4.68 	\\
9	&	LNS	&	1.15 	&	0.81 	&	1.90 	&	
1.49 	&	6.64 	&	5.47 	\\
10	&	SK255	&	0.89 	&	0.62 	&	1.55 	&	
1.17 	&	5.37 	&	4.40 	\\
11	&	BSK17	&	1.02 	&	0.70 	&	1.54 	&	
1.23 	&	5.68 	&	4.75 	\\
12	&	SAMi0	&	1.01 	&	0.74 	&	1.73 	&	
1.36 	&	6.13 	&	5.15 	\\
 \hline \hline
\end{tabular}
\end{table}

\begin{table}[t!]
\caption{The Coulomb mixing (\%) of $N = Z$ nuclei including $^{40}$Ca, 
$^{56}$Ni, and $^{100}$Sn. The Coulomb potential is \textit{excluded} in the 
HF-RPA calculation.
\label{IsomixNeqZ2}}
 
\centering
\begin{tabular}{|r|l|rr|rr|rr|}
 \hline \hline
& & \multicolumn{2}{c|}{$^{40}$Ca} 
& \multicolumn{2}{c|}{$^{56}$Ni} & 
\multicolumn{2}{c|}{$^{100}$Sn} \\
No. & Int. & $Q^{(\rm IV)}_0$ & $V_C^{(\rm IV)}$ 
& $Q^{(\rm IV)}_0$ & $V_C^{(\rm IV)}$ & 
$Q^{(\rm IV)}_0$ & $V_C^{(\rm IV)}$ \\
\hline
1	&	SIII	&	1.06 	&	0.80 	&	1.69 	&	
1.42 	&	5.54 	&	4.94 	\\
2	&	SGII	&	1.21 	&	0.94 	&	2.01 	&	
1.68 	&	6.70 	&	5.97 	\\
3	&	SKM*	&	1.23 	&	0.93 	&	1.97 	&	
1.64 	&	6.53 	&	5.76 	\\
4	&	SLy4	&	1.17 	&	0.92 	&	1.95 	&	
1.67 	&	6.40 	&	5.79 	\\
5	&	SAMi0	&	1.12 	&	0.88 	&	1.86 	&	
1.55 	&	6.21 	&	5.53 	\\
 \hline \hline
\end{tabular}
\end{table} 

In the case of nuclei that have $N > Z$, the Coulomb mixing and isospin mixing 
are different. Among $N > Z$ nuclei, $^{78}$Ni is an interesting nucleus 
because $T = 14$ is large while $Z = 20$ is relatively small. As one expects, 
the isospin mixing is strongly reduced by the factor $1/(T+1)$. In most nuclei, 
the Coulomb potential can be treated in perturbation theory. 
When $Z$ becomes large and the Coulomb potential becomes very strong 
such as in the case of Oganneson ($^{302}_{118}$Og), the perturbation theory 
in the calculation of the Coulomb mixing is not correct. However, the isospin 
mixing in $^{302}_{118}$Og ($T = 33$) is still small, around 2\%, because of 
the factor $1/(T+1)$. It is useful to remind that the uncertainties caused by 
other sources besides the Coulomb interaction are expected to be an order of 
magnitude smaller, thus they are smaller than the difference between results 
obtained with different choices of the Skyrme interactions.

\begin{table}[t!]
\caption{Coulomb mixing $\varepsilon^2_C$ (\%) and isospin mixing 
$\varepsilon^2_{T+1}$ (\%) of $N>Z$ nuclei including $^{48}$Ca ($T = 4$), 
$^{78}$Ni ($T = 14$), $^{90}$Zr ($T = 5$), $^{120}$Sn ($T = 10$), and 
$^{208}$Pb ($T = 22$).}
 
\centering
\begin{tabular}{|r|l|cc|cc|cc|cc|cc|}
 \hline \hline
 & & $^{48}$Ca & & $^{78}$Ni & & $^{90}$Zr &   & $^{120}$Sn &    & 
         $^{208}$Pb     & \\
        &               & $\varepsilon^2_C$ & $\varepsilon^2_{T+1}$ & 
$\varepsilon^2_C$ & $ \varepsilon^2_{T+1}$ &  $\varepsilon^2_C$ & 
$\varepsilon^2_{T+1}$ & $\varepsilon^2_C$ & $\varepsilon^2_{T+1}$ & 
$\varepsilon^2_C$ & $\varepsilon^2_{T+1}$\\
\hline
1 & SIII & 1.14  & 0.10  & 3.83  & 
0.04  & 4.13  & 0.52  & 10.85  & 0.23  & 
29.90  & 0.29  \\
2 & SGII & 1.33  & 0.12  & 4.28  & 
0.05  & 4.98  & 0.66  & 11.66  & 0.32  & 
34.38  & 0.41  \\
3 & SKM* & 1.35  & 0.12  & 4.66  & 
0.05  & 4.91  & 0.62  & 12.10  & 0.30  & 
35.57  & 0.36  \\
4 & SKP & 1.52  & 0.11  & 5.53  & 
0.04  & 5.18  & 0.58  & 12.41  & 0.28  & 
38.50  & 0.32  \\
5 & SkI2 & 1.20  & 0.09  & 4.71  & 
0.03  & 4.13  & 0.53  & 11.03  & 0.25  & 
32.36  & 0.32  \\
6 & SLy4 & 1.37  & 0.12  & 4.45  & 
0.04  & 4.73  & 0.56  & 11.36  & 0.25  & 
33.64  & 0.29  \\
7 & SKO & 1.32  & 0.05  & 6.19  & 
0.02  & 4.02  & 0.41  & 10.50  & 0.23  & 
32.75  & 0.26  \\
8 & SKO' & 1.20  & 0.07  & 4.79  & 
0.03  & 4.36  & 0.49  & 11.69  & 0.25  & 
34.64  & 0.28  \\
9 & LNS & 1.43  & 0.12  & 4.85  & 
0.04  & 5.10  & 0.63  & 12.86  & 0.29  & 
38.04  & 0.34  \\
10 & SK255 & 1.24  & 0.08  & 4.64  & 
0.03  & 4.13  & 0.49  & 10.84  & 0.24  & 
31.94  & 0.28  \\
11 & BSk17 & 1.18  & 0.10  & 4.40  & 
0.04  & 4.30  & 0.53  & 11.38  & 0.23  & 
33.24  & 0.28  \\
12 & SAMi0 & 1.35  & 0.11  & 4.27  & 
0.05  & 4.72  & 0.60  & 10.93  & 0.30  & 
31.85  & 0.37  \\
 \hline \hline
\end{tabular}
\end{table}

Finally, as mentioned in the text, for the isospin mixing, only the $T + 1$ 
states are taken into account and the isospin properties of the IVM resonance 
are useful for the calculation. Therefore, the properties of the IVM resonance 
are shown in Tables~\ref{48Catab1}-\ref{208Pbtab1} for $^{48}$Ca, $^{78}$Ni, 
$^{90}$Zr, $^{120}$Sn, and $^{208}$Pb, respectively.

In Tables~\ref{48Catab1}-\ref{208Pbtab1}, the average energy of the transition 
strength distribution is $\overline{E}_0 = m_0(1)/m_0(0)$. 
$S_T$ and $S_{T+1}$ are the total transition strength to the $|T, T; \alpha 
\rangle$ and $|T+1, T; \alpha \rangle$, respectively. $\Delta E_C$ is 
the direct CDE. $\Delta \overline{E}_+$ given by Eq.~(\ref{DeltaEplus}) is the 
difference in energy between $|T, T; \alpha \rangle$ and $|T+1, T; \alpha 
\rangle$, and $V_1$ is the symmetry potential defined in Eq.~(\ref{potV1}). 
These values can be compared to the work in Ref.~\cite{AUERBACH198377} where 
the Green function method was employed using the SIII Skyrme interaction.

\begin{table}[b!]
\caption{Isospin properties of the IVM resonance for $^{48}$Ca ($\hbar \omega = 
11.28$ MeV). $\overline{E}_0$ is the average energy of the strength 
distribution. $S_T$ and $S_{T+1}$ are the reduced transition strength to $|T, 
T; \alpha \rangle$,  and $|T+1, T; \alpha\rangle$, respectively. $\Delta E_C$ 
is the direct term of the CDE. $\Delta \overline{E}_+$ is the energy difference 
between $|T, T; \alpha \rangle$ and $|T+1, T; \alpha \rangle$. $V_1$ is the 
symmetry potential as defined in the expression (\ref{potV1}). \label{48Catab1}}
 
\centering
\begin{tabular}{|r|l|r|r|r|r|r|r|}
\hline
\hline
 &  & $\overline{E}_0$ & $S_{T}$ & $S_{T+1}$ & $\Delta E_C$ &  
$\Delta \overline{E}_+$ & 
$V_1$ \\
\hline
1 & SIII & 34.79 & 149.27 & 125.68 & 
7.27 & 11.06 & 106.14 \\
2 & SGII & 32.87 & 153.12 & 113.98 & 
7.33 & 6.88 & 66.01 \\
3 & SKM* & 32.54 & 156.72 & 125.68 & 
7.23 & 9.60 & 92.17 \\
4 & SKP & 30.00 & 160.48 & 123.00 & 
7.23 & 11.15 & 107.02 \\
5 & SkI2 & 33.08 & 141.64 & 91.56 & 
7.10 & 7.43 & 71.34 \\
6 & SLy4 & 30.57 & 148.43 & 123.51 & 
7.22 & 10.19 & 97.80 \\
7 & SKO & 32.60 & 144.58 & 72.78 & 
6.96 & 15.84 & 152.08 \\
8 & SKO' & 32.32 & 138.67 & 81.67 & 
7.14 & 11.75 & 112.77 \\
9 & LNS & 33.14 & 134.72 & 102.85 & 
7.52 & 9.78 & 93.86 \\
10 & SK255 & 34.65 & 153.56 & 100.39 & 
7.08 & 11.77 & 112.99 \\
11 & BSk17 & 32.83 & 139.00 & 110.69 & 
7.31 & 11.32 & 108.63 \\
12 & SAMi0 & 32.28 & 154.64 & 109.01 & 
7.20 & 8.05 & 77.25 \\
\hline
\hline
\end{tabular}
\end{table}

\begin{table}[t!]
\caption{The same as in Table~\ref{48Catab1}, but for $^{78}$Ni ($\hbar \omega 
= 9.60$ MeV). \label{78Nitab1}}
 
\centering
\begin{tabular}{|r|l|r|r|r|r|r|r|}
\hline
\hline
& & $\overline{E}_0$ & $S_{T}$ & $S_{T+1}$ & $\Delta E_C$ &  $\Delta 
\overline{E}_+$ & 
$V_1$ \\
\hline
1 & SIII & 30.55 & 382.04 & 186.47 & 
8.90 & 18.19 & 118.23 \\
2 & SGII & 28.86 & 385.18 & 186.63 & 
8.96 & 14.34 & 93.22 \\
3 & SKM* & 28.47 & 400.08 & 175.45 & 
8.85 & 16.29 & 105.91 \\
4 & SKP & 26.14 & 405.03 & 165.44 & 
8.88 & 18.40 & 119.57 \\
5 & SkI2 & 27.01 & 387.49 & 123.12 & 
8.64 & 19.35 & 125.79 \\
6 & SLy4 & 27.10 & 373.59 & 168.29 & 
8.87 & 17.50 & 113.74 \\
7 & SKO & 25.51 & 412.48 & 117.36 & 
8.60 & 26.73 & 173.73 \\
8 & SKO' & 26.77 & 365.85 & 112.93 & 
8.80 & 21.14 & 137.41 \\
9 & LNS & 29.24 & 341.93 & 121.66 & 
9.18 & 17.14 & 111.43 \\
10 & SK255 & 29.37 & 401.49 & 120.28 & 
8.64 & 19.59 & 127.33 \\
11 & BSk17 & 28.01 & 364.18 & 155.15 & 
8.93 & 19.89 & 129.29 \\
12 & SAMi0 & 28.96 & 388.83 & 196.12 & 
8.87 & 15.02 & 97.66 \\
\hline
\hline
\end{tabular}
\end{table}

\begin{table}[b!]
\caption{The same as in Table~\ref{48Catab1}, but for $^{90}$Zr ($\hbar \omega 
= 9.15$ MeV). \label{90Zrtab1}}
 
\centering
\begin{tabular}{|r|l|r|r|r|r|r|r|}
\hline
\hline
& & $\overline{E}_0$ & $S_{T}$ & $S_{T+1}$ & $\Delta E_C$ &  $\Delta 
\overline{E}_+$ & 
$V_1$ \\
\hline
1 & SIII & 33.53 & 425.52 & 430.42 & 
12.22 & 4.24 & 63.66 \\
2 & SGII & 31.45 & 426.10 & 444.33 & 
12.38 & 3.52 & 52.74 \\
3 & SKM* & 31.53 & 435.10 & 445.68 & 
12.25 & 3.98 & 59.64 \\
4 & SKP & 29.61 & 431.82 & 424.44 & 
12.28 & 5.30 & 79.56 \\
5 & SkI2 & 31.89 & 380.95 & 368.23 & 
12.04 & 3.12 & 46.82 \\
6 & SLy4 & 29.96 & 405.62 & 405.58 & 
12.25 & 4.90 & 73.44 \\
7 & SKO & 32.08 & 377.30 & 293.82 & 
11.85 & 3.41 & 51.18 \\
8 & SKO' & 30.98 & 375.09 & 330.47 & 
12.08 & 3.71 & 55.64 \\
9 & LNS & 32.08 & 371.34 & 385.09 & 
12.70 & 4.82 & 72.28 \\
10 & SK255 & 33.93 & 412.47 & 400.59 & 
12.01 & 4.60 & 69.04 \\
11 & BSk17 & 31.61 & 387.62 & 380.95 & 
12.34 & 4.32 & 64.73 \\
12 & SAMi0 & 31.78 & 424.63 & 426.62 & 
12.23 & 3.59 & 53.78 \\
\hline
\hline
\end{tabular}
\end{table}

\begin{table}[b!]
\caption{The same as in Table~\ref{48Catab1}, but for $^{120}$Sn ($\hbar \omega 
= 8.31$ MeV). \label{120Sntab1}}
 
\centering
\begin{tabular}{|r|l|r|r|r|r|r|r|}
\hline
\hline
& & $\overline{E}_0$ & $S_{T}$ & $S_{T+1}$ & $\Delta E_C$ &  $\Delta 
\overline{E}_+$ & 
$V_1$ \\
\hline
1 & SIII & 29.86 & 804.40 & 469.01 & 
13.97 & 8.73 & 95.20 \\
2 & SGII & 28.25 & 798.77 & 499.43 & 
14.12 & 6.87 & 74.98 \\
3 & SKM* & 28.21 & 818.03 & 492.10 & 
13.92 & 7.51 & 81.94 \\
4 & SKP & 26.77 & 795.35 & 466.98 & 
14.00 & 8.39 & 91.48 \\
5 & SkI2 & 27.78 & 735.27 & 402.59 & 
13.73 & 8.03 & 87.55 \\
6 & SLy4 & 26.96 & 756.12 & 427.30 & 
13.97 & 8.91 & 97.18 \\
7 & SKO & 28.10 & 695.72 & 401.01 & 
13.95 & 8.53 & 93.08 \\
8 & SKO' & 27.04 & 709.41 & 405.85 & 
13.99 & 8.98 & 97.98 \\
9 & LNS & 28.70 & 701.41 & 414.99 & 
14.37 & 8.49 & 92.57 \\
10 & SK255 & 30.11 & 777.41 & 453.56 & 
13.67 & 8.43 & 91.93 \\
11 & BSk17 & 27.87 & 740.77 & 404.41 & 
14.03 & 9.12 & 99.45 \\
12 & SAMi0 & 28.70 & 788.05 & 490.95 & 
14.08 & 7.21 & 78.63 \\
\hline
\hline
\end{tabular}
\end{table}

\begin{table}[b!]
\caption{The same as in Table~\ref{48Catab1}, but for $^{208}$Pb ($\hbar \omega 
= 6.92$ MeV). \label{208Pbtab1}}
 
\centering
\begin{tabular}{|r|l|r|r|r|r|r|r|}
\hline
\hline
& & $\overline{E}_0$ & $S_{T}$ & $S_{T+1}$ & $\Delta E_C$ &  $\Delta 
\overline{E}_+$ & 
$V_1$ \\
\hline
1 & SIII & 28.19 & 2081.68 & 1164.34 & 
19.34 & 9.87 & 89.25 \\
2 & SGII & 26.35 & 2104.85 & 1271.02 & 
19.52 & 7.90 & 71.45 \\
3 & SKM* & 26.50 & 2130.51 & 1221.69 & 
19.36 & 8.81 & 79.69 \\
4 & SKP & 25.04 & 2065.16 & 1111.53 & 
19.46 & 10.59 & 95.75 \\
5 & SkI2 & 25.37 & 1972.33 & 1012.50 & 
19.00 & 8.64 & 78.12 \\
6 & SLy4 & 25.29 & 1954.69 & 1025.64 & 
19.39 & 10.85 & 98.12 \\
7 & SKO & 25.28 & 1918.29 & 895.07 & 
19.08 & 9.28 & 83.90 \\
8 & SKO' & 24.94 & 1878.92 & 928.52 & 
19.28 & 10.06 & 91.01 \\
9 & LNS & 26.85 & 1839.60 & 1034.58 & 
19.93 & 10.29 & 93.03 \\
10 & SK255 & 27.89 & 2064.87 & 1106.52 & 
18.98 & 10.00 & 90.46 \\
11 & BSk17 & 26.04 & 1918.62 & 986.63 & 
19.50 & 10.48 & 94.80 \\
12 & SAMi0 & 26.87 & 2077.67 & 1215.95 & 
19.45 & 8.10 & 73.26 \\
\hline
\hline
\end{tabular}
\end{table}

In Tables~\ref{48Catab1}-\ref{208Pbtab1}, the values of $\hbar \omega = 41 
\times A^{-1/3}$ are given to describe the $A^{-1/3}$ behavior of the energy of 
the IVM- $\overline{E}_0$, from our calculation. This behavior was also 
obtained in the hydrodynamical model or other collective models. 
Due to the Pauli blocking of the excess neutrons, it is expected 
that $S_{T+1}/S_{T} < 1$. Indeed, the p-h excitations involving a 
proton transformed into a neutron in an excess neutron orbit are forbidden by 
the Pauli principle. Not so in the excitations represented by $S_T$. There one 
can have p-h components in which the proton is placed in the orbits 
occupied by the excess neutrons. However, it is different in the case of 
$^{90}$Zr. A plausible explanation is in the following. In $^{90}$Zr the excess 
neutrons occupy mostly the $1g_{9/2}$ orbit. There are no $j = 9/2$ orbit below 
the $1g_{9/2}$ and therefore one cannot occupy this orbit if one constructs a 
$J^\pi = 0^+$ state. Thus the amount of p-h components for the $T$, 
and $T+1$ is the same and there should be little difference between $S_T$ and 
$S_{T+1}$. In $\Delta E_C$, only the direct part which contributes more than 
90\% to the CDE was included. Other effects: exchange term, finite proton size 
effect, and vacuum polarization \cite{RevModPhys.44.48} were not taken into 
account. All these corrections are small and contribute only a few percents to 
the CDE. It makes the value of the \textit{direct} CDE quite acceptable for the 
purpose of our study. The value of $V_1$ obtained from a 
single-particle-symmetry potential is about 100 MeV \cite{BohrMottelsonI}. In 
our calculation, $V_1$ is also around 100 MeV.

\section{Conclusion}
The isospin mixing in the ground state is small, especially in heavy nuclei. It 
does not strongly depend on the choice of the Skyrme interactions. It is clear 
that the formalism of isospin is not only useful but also powerful in nuclear 
physics, where many examples of isospin symmetry can be found. In particular, 
this is true in the ground states where isospin mixing does not exceed a few 
percents. This does not mean that the isospin non-conserving interaction, the 
Coulomb force plays a minor role in forming the nucleus, as can be seen in the 
large Coulomb mixing we calculated.

\begin{acknowledgments}
The authors thank to Nguyen Van Giai for discussions and Vladimir Zelevinsky 
for discussions made possible by the travel grant from the US-Israel Binational 
Science Foundation (2014.24). This work was supported by the US-Israel 
Binational Science Foundation, grant 2014.24.
\end{acknowledgments}

\pagebreak

\end{document}